\documentclass[12pt]{article}
\usepackage{graphicx}

\begin{document}

\title
{Transitional dynamics of a phase qubit-resonator system:
 requirements for fast readout of a phase qubit}
\author{G.P. Berman $^{1}$\footnote{Corresponding
author: gpb@lanl.gov}, A.A. Chumak$^{1,2}$, and V.I. Tsifrinovich$^{3}$
\\[3mm]$^1$
 Theoretical Division, MS-B213, Los Alamos National Laboratory, \\ Los Alamos, NM 87545
\\[5mm] $^2$  Institute of Physics of
the National Academy of Sciences,\\ pr. Nauki 46, Kiev-28, MSP 03028,
Ukraine
\\[3mm]$^3$
 Department of Applied Physics, Polytechnic Institute of NYU,\\ 6 MetroTech Center, Brooklyn, NY 11201
\\[3mm]\rule{0cm}{1pt}}
\maketitle \markright{right_head}{LA-UR-11-11790}

\begin{abstract}

We examine the non-stationary evolution of a coupled qubit-\\transmission line-resonator system coupled to  an
external drive and the resonator environment.
By solving the equation for a non-stationary resonator field, we determined the requirements for a
single-shot non-destructive dispersive measurement of the phase qubit state. Reliable
isolation of the qubit from the ``electromagnetic environment" is
necessary for a dispersive readout and can be achieved if the whole
system interacts with the external fields only through the resonator that is
weakly coupled to the qubit. A set of inequalities involving
resonator-qubit detuning and coupling parameter, the resonator leakage
and the measurement time, together with the requirement of multi-photon
outgoing flux is derived. It is shown, in particular, that a decrease of the
measurement time requires an increase of the resonator leakage. This increase
 results in reducing the quality factor and decreasing the resolution
of the resonator eigenfrequencies corresponding to different
qubit states. The consistent character of the derived inequalities
for two sets of experimental parameters is discussed.

The obtained results will be useful for optimal design of experimental setups, parameters,
and measurement protocols.

\end{abstract}

\section{Introduction}

The measurement of qubits is an important step in quantum
information processing.
For accurate qubit detection, the readout has to be faster than the
qubit relaxation time. Moreover, to implement quantum error
correction, the readout time must be less than the decoherence time.
The relaxation processes can be of different origin (even due to the
qubit-vacuum interaction). Also, the effects of measuring devices on
the behavior of qubit states during the readout time are of great importance
\cite{el1}-\cite{el3}.

The simplest scheme for fast qubit readout of the phase qubit, based on the tunneling effect, (single-shot measurement) has been implemented and studied in
\cite{coo}-\cite{kof}. According to this scheme, the measurement pulse
adiabatically reduces the barrier between the potential wells, one of which forms the qubit states. As a result, the qubit in the upper state
switches by tunneling into the neighboring well with probability close
to one, while the qubit in the lower state remains unchanged. This kind
of readout is limited by the strong current noise back action from the
measurement device on the qubit. Moreover, this demolition measurement process
destroys the qubit states which is unacceptable for most applications.

The scheme of dispersive readout of qubit states  has evident advantages.
In what follows, we consider the case in which an external device
probes the qubit state indirectly via a transmission line
resonator (a two sided cavity) weakly coupled to the qubit. This
scheme allows repeated measurements. The
presence of the qubit-resonator coupling causes the resonator
eigenfrequency to be dependent on the qubit state. Hence, the number
of photons in the resonator, being dependent on the probe
field-resonator detuning, depends on the qubit state. (This number
is maximal for zero detuning.) For this reason, the field going out of the resonator
contains information about the resonator state as well
as about the qubit state.

The amplitude of the transmitted field depends on the photon number
in the resonator and on the resonator leakage. The increase of both
quantities results in an increase of the transmitted signal that in
general improves the signal-to-noise ratio. However, both of these
factors decrease the qubit relaxation time due to the qubit-resonator interaction.
Moreover, resonators with high leakage (low quality factor) lose their
ability to filter out those external noises which penetrate into
the low-temperature region from the ``external world". Also, for a relatively large number of photons in the resonator, the
qubit-resonator interaction cannot be considered to be weak, and the
advantages of the dispersive measurement, based on the perturbation approach, disappear.

The optimal choice of both the readout strategy and the setup parameters can be
facilitated by means of theoretical description of the physical
processes involved in the course of readout. In what follows, we use our recent
approach \cite{ber} which makes it possible to
obtain explicit expressions for the photon numbers and the measurement-induced
relaxation rate of the qubit. We derive a set of inequalities which are required to provide a single-shot dispersive measurement, and we present examples of the qubit and resonator parameters.

\section{Hamiltonian and equations of motion}

As in our previous paper \cite{ber}, we consider a flux-biased phase
qubit embedded in a symmetric transmission line by a set of
distributed inductances and capacitances. This is shown
schematically in Fig. 1. The qubit is coupled to the center of the
transmission line by the capacitance, $C_g$, leading to their
effective interaction strength, $g$. The qubit transition frequency,
$\omega_q$, is tunable by varying an external flux, $\Phi$, shown
schematically in Fig. 1. It is assumed that $\omega_q$ as well as
the frequency of the external drive, $\omega_d$, is close to one of
the eigenfrequencies of the transmission line, $\omega_r$. In this
case, one can consider the transmission line as a resonator with a
lumped inductance, $L_r$, and a capacitance, $C_r$. Then, the
transmission line can be described in terms of a harmonic oscillator
characterized by the single eigenfrequency,
$\omega_r=(L_rC_r)^{-1/2}$.

The total Hamiltonian is,
\[ H=-{1\over 2}\hbar\omega_q\sigma_z +\hbar\omega_r\bigg
(a^\dag a+{1\over2}\bigg )+\sum_n \hbar\omega_n\bigg(b^\dag
_nb_n+{1\over 2}\bigg )+\]
\begin{equation}\label{one}
i\hbar g\sigma_y\bigg (a^\dag-a\bigg)+i\hbar \sum_n f_n\bigg
(b_na^\dag -b^+_na\bigg )+i\hbar f_0\bigg (ca^\dag -c^\dag a\bigg ),
\end{equation}
where $\sigma_z$ is the Pauli operator, $a^\dag ,b_n^\dag ,c^\dag
(a,b_n,c)$ are the creation (annihilation) operators of the
resonator, bath, and drive excitations, respectively. The bath is
modeled by an infinite set of harmonic oscillators. (See Refs.
\cite{wal} and \cite{int}.)

   The qubit-resonator interaction term (the fourth term
in (\ref{one})) couples the variables of both subsystems. It is
evident that only even harmonics of the transmission line are
coupled with the qubit. We consider that the second harmonic
has a frequency close to the qubit frequency. In the center of the transmission
line, the
voltage of the second harmonic is equal to \cite{el1},
 \[V(t)=-\sqrt{\frac {\hbar \omega_r}{Lc}}\big (a+a^+\big),\]
where $L$ and $c$ are, correspondingly, the length of the line and
its capacitance per unit length. For a given term for the voltage,
$V(t)$, applied to both, $C_g$ and the qubit loop, we can easily
express the interaction strength, $g$, as a function of other
parameters of the system. This interaction strength, $g$, is given
by the expression,
\[g=C_g\sqrt{\frac{\omega_r\omega_q}{2LcC_J}}.\]
For calculational simplicity, we approximated the potential
energy of the qubit by a parabolic dependence on the effective coordinate.
More details can be found in \cite{arx}.

As one can see from Eq. \ref{one}, the interaction term is linear in the qubit
and the resonator variables. Alternative interaction Hamiltonians
(nonlinear in the resonator variables) describe the situation
considered in \cite{lup} - \cite{sid}.
\begin{figure} [ht]
\centering
\includegraphics{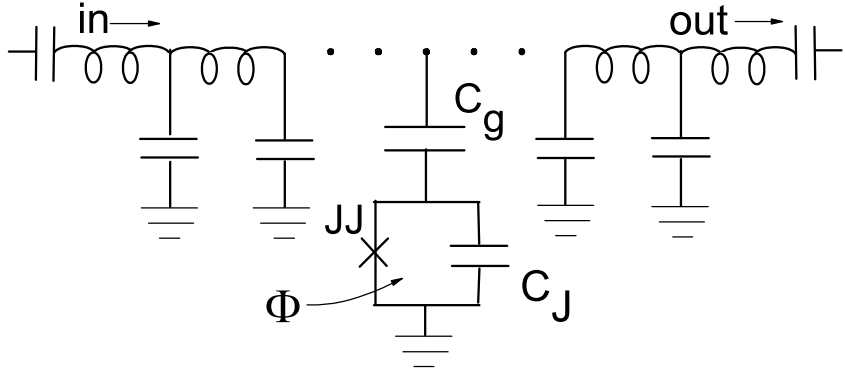}
\caption{A scheme of the qubit-transmission-line system. $C_J$ is
the capacitance shunting the Josephson junction, JJ. $C_g$ is the
capacitance coupling the qubit and the transmission line. $\Phi$ is
the external flux threading the qubit loop.}
\end{figure}
The resonator-thermostat and resonator-drive interaction terms are similar to the resonator-qubit interaction. The analytical approach
in Ref. \cite{ber} is based on a set of inequalities,
\begin{equation}\label{two}
|\omega_{ij}|<<\omega_i,\omega_j,\quad
\omega_{ij}\equiv\omega_i-\omega_j,
\end{equation}
where the indices $i,j$ indicate the corresponding frequencies,
$\omega_{q,r,d}$. The relations (\ref{two}) indicate that all frequencies
are close to each other.

Moreover, inequalities (\ref{two}) were complemented by the requirement of a
dispersive regime for our Hamiltonian. Usually, this is written in a
form,
\begin{equation}\label{thr}
 \frac g{|\omega_{qr}|}<<1.
 \end{equation}
Taking into account the fact that the interaction term depends linearly on
$a$ and $a^\dag$, a more accurate criterion can be written:
\begin{equation}\label{fou}
 \frac {g\sqrt{\tilde{ n}_r}}{|\omega_{qr}|}<<1,
 \end{equation}
where $\tilde{ n}_r$ is the characteristic photon number in the
resonator.

Considering all of the operators in the Heisenberg representation, we have
derived in \cite{ber} the equation of motion for the resonator field,
$a(t)$,
\begin{equation}\label{fiv}
\bigg [\partial _t+i\bigg (\omega_r- \chi\sigma_z\bigg )+
\frac{\kappa}{2}\bigg ]a=f_0c+\sum_nf_n\tilde{b}_n-
ig\tilde{\sigma}_+,
\end{equation}
where  $\chi \equiv {g^2}/\omega_{qr}$, and the operators in the
right-hand side have the following time-dependence:
\[c(t)=c(t_0)e^{-i\omega_d(t-t_0)},\quad \tilde{b}_n(t)=b_n(t_0)
e^{-i\omega_n(t-t_0)},\quad
\tilde{\sigma}_+(t)=\sigma_+(t_0)e^{-i\omega_q(t-t_0)}.\] The last
term in the square brackets comes from the bath dynamics, and
represents the linear damping of the cavity mode. The right-hand side of
Eq. (\ref{fiv}), in which operators evolve as if there is no
interaction with the cavity, can be interpreted as the input mode.
One can visualize that assuming that in the remote past (at
$t_0,\, t>t_0$) the corresponding excitations were moving
towards the cavity but were not still affected by it.

In deriving Eq. (\ref{fiv}), we used Eqs. (\ref{two})
and (\ref{fou}). Also, $\sigma_z$ was assumed as a slowly varying
quantity. The consideration below follows the general scheme outlined in
\cite{wal} and \cite{int}.

Ignoring the initial conditions and the transient stage, the solution
of Eq. (\ref{fiv}) can be written as:
\begin{equation}\label{six}
 a(t)=\frac {if_0c(t)}{\tilde{\omega}_{dr}+i\kappa/2}+
 \sum_n\frac
 {if_n\tilde{b}_n(t)}{\tilde{\omega}_{nr}+i\kappa/2}+
 \frac {g\tilde{\sigma}_+(t)}{\tilde{\omega}_{qr}+i\kappa/2},
\end{equation}
where $\tilde{\omega}_{ir}=\omega_i-\omega_r+\chi \sigma_z$.

Strictly speaking, there cannot be steady state of the system with
the qubit in the excited state. Nevertheless, there are long-lived
quasi-steady states with given values of $\sigma_z=\pm 1$ if the qubit
relaxation time is much greater than that of the resonator (i.e.
$\kappa^{-1}$). To understand at which parameters and time intervals
Eq. (\ref{six}) can adequately describe the state of the cavity field,
an evolution equation for $\sigma_z$ is required. Within the
dynamics governed by the Hamiltonian (\ref{one}), it follows from
\cite{ber} that,
\[\frac \partial{\partial t}{\sigma}_z=\frac {2g^2}{\omega_{qr}^2}\bigg\{\bigg [2
\omega_{qr}\sin(\omega_{qr}t)e^{-\frac {\kappa} 2t}+\kappa
\bigg(1-\cos(\omega_{qr}t)e^{-\frac \kappa 2t}\bigg)\bigg ]
\bigg[\frac 12-\sigma_z\bigg(n_r^b+\frac 12\bigg) \bigg]-
\]
\begin{equation}\label{sev}
\sigma_z\frac {|f_0c|^2}{\tilde{\omega}^2_{dr}+\kappa^2/4}\bigg[2
\tilde{\omega} _{dr}\bigg(1-\cos(\omega_{qr}t)e^{-\frac \kappa
2t}\bigg)+\kappa \sin(\omega_{qr}t)e^{-\frac \kappa 2t}\bigg] \bigg
\},
\end{equation}
where $n_r^b$ is the number of photons in the resonator, averaged
over bath variables. It is assumed that other relaxation mechanisms
are not as important as those included in Eq. (\ref{sev}).

Considering only the initial stage  of the excited state relaxation
under resonant conditions ($\tilde{\omega} _{dr}=0$), we can omit
the second term in the braces of (\ref{sev}). Then, Eq. (\ref{sev})
reduces to,
\begin{equation}\label{eig}
\frac \partial{\partial t}\bigg\{{\sigma}_z+\frac
{2\chi}{\omega_{qr}}\cos(\omega_{qr}t)e^{-\frac {\kappa} 2t}\bigg[
1-\sigma_z\bigg(2n_r^b+1\bigg) \bigg]\bigg\}=\frac
{\chi}{\omega_{qr}} \kappa \bigg[1-{\sigma}_z\bigg(2n_r^b+1\bigg)
\bigg],
\end{equation}
where the condition, $|\omega_{qr}|>>\kappa/2$, consistent with the
dispersive regime, was assumed. By neglecting the right-side
term, we obtain an equation describing the small-amplitude Rabi
oscillations with frequency, $\omega_{qr}$, in the vicinity of
$\sigma_z=-1$. Its solution is given by:
\begin{equation}\label{nin}
\sigma_z(t)\approx -1+\frac {4\chi}{\omega_{qr}}\bigg
(1-cos(\omega_{qr}t)e^{-\frac {\kappa} 2t}\bigg
)\bigg(n_r^b+1\bigg).
\end{equation}
 The oscillations decay for times greater than $2/\kappa$.
 Their  influence on the measurement fidelity was
 analyzed in \cite{ber}.

 The total relaxation of the qubit can be described by taking into account
 the right-hand side of Eq. (\ref{eig}). It can be easily seen that the
 relaxation rate, $\gamma_r$, is given by,
\begin{equation}\label{ten}
\gamma_r=\frac {2\chi}{\omega_{qr}} \kappa \bigg(n_r^b+1\bigg).
\end{equation}
As we see, the qubit relaxation occurs even in the case of empty
resonator ($n_r^b=0$) due to the resonator leakage,
$\kappa$. In the literature, this effect is known as a vacuum-induced
relaxation.

The condition for nondestructive measurement of the excited state can
be expressed in the form of the inequality,
\begin{equation}\label{ele}
\gamma_r(t-t_0)<<1,
\end{equation}
where, $t-t_0$, is the measurement time.

\section{Non-steady state of the resonator field}

Due to the measurement, the resonator state can vary with
time. Following Ref. \cite{bia}, we consider the dynamics of
a resonator that was initially in the steady state corresponding
to the qubit in the ground state. Only an insignificant number of
photons is in  the resonator because of the cavity-drive is detuned
by $-2\chi$. It is assumed that at $t=t_0$, a strong $\pi$-pulse is
applied directly to the qubit at its transition frequency. After
this, the cavity-drive resonant conditions are satisfied
($\omega_d=\omega_r+\chi$), and the field amplitude, $a(t)$, starts to
grow. Using the approach of \cite{ber}, we can analytically obtain
an expression for the resonator field. With the initial (at $t=t_0$) value of $a(t)$
given by steady-state term, Eq. (\ref{six}), we can easily solve  Eq.
(\ref{fiv}),
\[a(t)=e^{(-i\omega_d-\kappa
/2)(t-t_0)}a(t_0)+\int_{t_o}^{t}dt^\prime e^{(-i\omega_d-\kappa
/2)(t-t^\prime)}\times\]
\begin{equation}\label{tve}
\bigg[f_0c(t^\prime)+\sum_nf_n\tilde{b}_n(t^\prime)-
ig\tilde{\sigma}_+(t^\prime)\bigg ].
\end{equation}
Retaining only the first term in square brackets of Eq. (\ref{tve}) and
integrating over $t^\prime$, we get:
\begin{equation}\label{Thi}
a(t)=f_0c(t)\bigg[\frac
{e^{-\kappa(t-t_0)/2}}{-i2\chi+\kappa/2}+\frac
{1-e^{-\kappa(t-t_0)/2}}{\kappa/2}\bigg].
\end{equation}
The first term in square brackets of Eq. (\ref{Thi}) is due to the
``memory" of the ``steady state". The second term describes
the increase of the resonator field due to the qubit excitation. In the
absence of the qubit transition, the resonator field would be:
\begin{equation}\label{Fou}
a(t)=\frac {f_0c(t)}{-i2\chi+\kappa/2}.
\end{equation}

It can be seen from Eqs. (\ref{Thi}) and (\ref{Fou}) that the
difference in the resonator fields corresponding to different qubit
states vanishes for short measurement times, $t-t_0<<<2/\kappa$. To
notice the difference in outgoing signals,  the conditions,
\begin{equation}\label{ad1}
\kappa/2 \geq(t-t_0)^{-1},
\end{equation}
as well as,
\begin{equation}\label{ad2}
2\chi \geq\kappa/2
\end{equation}
must be satisfied.

It follows from Eqs. (\ref{ad1}) and (\ref{ad2}) that $2\chi\geq
(t-t_0)^{-1}$. This inequality can also be derived from the
following general consideration.  The spectral width of any
$(t-t_0)$-long pulse is not smaller than, $(t-t_0)^{-1}$. Therefore
in the opposite case, $2\chi<<(t-t_0)^{-1}$, two harmonics with
frequencies, $\omega_r\pm\chi$, being close to one another within this
spectral interval, have approximately equal amplitudes. So, these harmonics cannot be easily resolved.

There is an essential difference in the values given
by Eqs. (\ref{Thi}) and (\ref{Fou}), if $t-t_0>2/\kappa$. In this case,
the value of outgoing field, $b_{out}(t)$, provides useful
information about the qubit state. Using input-output theory
\cite{wal}-\cite{int} and considering the resonator as a symmetric
two-sided cavity, the transmitted field is determined by the
resonator field as,
\begin{equation}\label{Fif}
b_{out}(t)=-\sqrt{\kappa\over2}a(t).
\end{equation}
The average photon number leaving the resonator through the right
side per unit time, is given by,
\begin{equation}\label{Six}
\langle b^+_{out}(t) b_{out}(t)\rangle={\kappa\over 2}\langle
a^+(t)a(t)\rangle.
\end{equation}
Substituting $a(t)$ from Eqs. (\ref{Thi}) or (\ref{Fou}) and considering
$c(t)$ as a classical variable, we can easily calculate the outgoing
photon flux.

The total number of photons, $n_T$, leaving the resonator during the
measurement time, $t-t_0$, is given by,
\begin{equation}\label{Sev}
n_T={\kappa\over 2}(t-t_0)\bar{n}\equiv{\kappa\over 2}\int_{t_0}^t
dt^\prime\langle a^+(t^\prime)a(t^\prime)\rangle.
\end{equation}
It can be seen that by definition, $\bar{n}$ is the photon number in
the resonator averaged over the measurement time, $t-t_0$. The required
 value of $\bar{n}$ can be achieved by varying the drive.
For theoretical estimates, Eqs. (\ref{Thi}), (\ref{Fou}), and (\ref{Sev})
can be used.

To get reliable information about the state of the qubit after the
$\pi$-pulse (if it is in the ground or excited state) in the course
of a single-shot measurement, many photons ($n_T>>1$) must interact
with the measurement device. It follows from Eq. (\ref{Sev}) that
this condition can easily be satisfied by increasing the measurement
time and (or) by increasing the leakage parameter, $\kappa$.
Unfortunately, both ways have evident disadvantages in view of
possible applications of qubits: (a) quantum computers require fast
measurements and (b) large values of $\kappa$ result in small
quality factors, $Q$, for the resonators ($Q=\omega_r/\kappa$). The
last circumstance can shorten the lifetime of the excited qubit
state.

The interplay of several factors should be taken into account for
the optimal choice of the measurement strategy. In the next Section, we
consider this issue in more details.

\section{Conditions required for a nondemolition single-shot dispersive measurement}

We start from the inequality (\ref{ele}), which provides a qubit to remain
in the excited state (with high probability, close to unity) after the
individual measurements. For the non-steady state, expression
(\ref{ele}) should be rewritten as:
\begin{equation}\label{Eig}
\frac {4g^2}{\omega_{qr}^2}{\kappa \over 2}(t-t_0)\bar{n}<<1,
\end{equation}
where $\bar{n}$ is defined by Eq. (\ref{Sev}). In Eq. (\ref{Eig}), we
have neglected  unity in comparison with the average photon
number, $\bar{n}$, in the resonator. Eq. (\ref{Eig}) can be rewritten
in the form,
\begin{equation}\label{Nin}
{\kappa \over 2}(t-t_0)\bar{n}<<\frac {\omega_{qr}^2}{4g^2},
\end{equation}
which is more convenient for physical interpretation. In particular,
the left-hand side term is the total photon number which left the cavity
during the  time, $t-t_0$. Eq. (\ref{Nin}) establishes an upper bound on this
number. At the same time, this number should be sufficiently large.
In this case, the measurement process acquires classical properties
in spite of the fact that a quantum state is measured.
Thus, we have,
\begin{equation}\label{Twe}
1<<{\kappa \over 2}(t-t_0)\bar{n}<<\frac {\omega_{qr}^2}{4g^2}=\frac
{\omega_{qr}}{4\chi} .
\end{equation}

It seems that the second inequality can be easily satisfied by
choosing a large  qubit-resonator detuning, $\omega_{qr}$, or a small
qubit-resonator interaction parameter, $g$. However, the variations of these
quantities are restricted by the requirement,
\begin{equation}\label{Two}
2|\chi|=2\frac {g^2}{|\omega_{qr}|}>\kappa/2,
\end{equation}
which means that nonresonant eigenfrequency should be beyond the
resonator bandwidth.

Inequalities (\ref{Twe}) and (\ref{Two}) should be complemented by two more
which were discussed in the previous Sections. The first inequality concerns
the validity of the dispersive approach, and follows directly from
Eq. (\ref{fou}). In the case of non-steady state, it is given by,
\begin{equation}\label{Ttw}
 \frac {g\sqrt{\bar {n}}}{|\omega_{qr}|}<<1.
 \end{equation}
The other inequality deals with the problem of the resolution of the two resonator
eigenfrequencies during the  measurement time,
\begin{equation}\label{Tth}
\kappa/2\geq(t-t_0)^{-1}.
\end{equation}
In summary, conditions (\ref{Twe})-(\ref{Tth}) should be satisfied for
nondestructive single-shot dispersive readout of the phase qubit.
Are these conditions non-contradictory? Is it possible to satisfy all of them,
 at least in some particular cases? In the next
Section we will illustrate such possibilities.

\section{Examples}

(i) We first choose the frequencies of the resonator and the qubit, and the coupling strength as: $\omega_r/2\pi=6.5\, GHz$, $\omega_q/2\pi=8
\,GHz$, and $g=0.05\omega_{qr}$. Then, the frequency shift is: $\chi/2\pi =3.75\, MHz.$ If
the measurement time is a given parameter, for example,
$t-t_0=4*10^{-8}\, s$, then the leakage value, $\kappa$, can be
taken as: $0.5*10^{8}\,s^{-1}$. This quantity is sufficiently large to
distinguish the contributions in $b_{out}$ arising from steady- and
non-steady states of the resonator (see Eq. (\ref{Thi})). At the same
time, it is not large enough to violate the conditions (\ref{Twe}) and
(\ref{Two}).

Finally, it is reasonable to assume $n_T=10$, hence, $\bar{n}=10$.
The value of $n_T$ is sufficiently larger than the fluctuations of
the outgoing photons (in the case of Poisson's statistics $\sqrt
{\langle{\delta n_T}^2\rangle}= \sqrt {n_T}$). On the other hand,
conditions (\ref{Twe}) and (\ref{Ttw}) are not violated in this case.

 The relaxation rate of the qubit due to the interaction with the resonator is: $\gamma_r=2.5*10^6\,s^{-1}$. Then,
 the probability of remaining in the excited state after the measurement, determined by
 $1-\gamma_r(t-t_0)$, is 0.9.

 The resonator quality factor is equal to  $Q=\omega_r/\kappa\approx 820$. The relatively small value of the quality factor is due to the large leakage from the resonator.  The power,
 coupled to the outside of the resonator, $P={\kappa \over
 2}\bar{n}\hbar\omega_r$,  determines the corresponding voltage, $\langle V_{out}\rangle$, via the relationship,
 $P=\langle V_{out}\rangle ^2/R$, where $R$ is the impedance at the
 output of the resonator. Usually $R=50\,\Omega$. Then, $\langle V_{out}\rangle
 =0.23 \,\mu V$.

 (ii) For comparison, let us consider
 resonator and the qubit frequencies that are a factor of two larger than in the case (i):
$\omega_r/2\pi=13\, GHz$, $\omega_q/2\pi=16 \,GHz$.

As before, $g=0.05\omega_{qr}$. Then $\chi/2\pi =7.5\, MHz.$
If the photon numbers and measurement time remain unchanged ($n_T=10$,
$\bar{n}=10$, $t-t_0=4*10^{-8}\, s$), then $\kappa
=0.5*10^{8}\,s^{-1}$. Because of the increase of $\omega_r$, the
quality factor as well as the ratio $\chi/\kappa$ doubles. A
similar tendency concerns the increase of the measurement time. For
greater $t-t_0$, the value of $\kappa $ can be smaller resulting in a larger
quality factor. Also, in this case, the output voltage, $\langle V_{out}\rangle$,   increases by a factor of $\sqrt{2}$.

 The increase of
the frequencies makes it easier to choose the desirable parameters required for a single-shot dispersive measurement.
For example, it becomes possible to decrease the relaxation
rate, $\gamma_r$, improving in this way the measurement fidelity. To show
this, let us assume in the case (ii) that $g/\omega_{qr}=0.05/{\sqrt 2}$.
Then, the relaxation time of the excited state doubles,
and the measurement fidelity grows from $90\% $ to $95\% $. In spite
of the decrease of $\chi/\kappa $ toward the value calculated for the
case (i), the resonator lines remain well-resolved. It is important to
emphasize, that a similar variation of $g/\omega_{qr}$ in the case
(i) will result in a very small $\chi$ being insufficient for resolving the lines.

\section{Conclusion}

At first sight, the statement that a single-shot dispersive measurement of the
phase qubit can also be nondemolition measurement appears unrealistic. A simple analysis of inequalities (\ref{Twe})-(\ref{Tth})
shows that it is rather difficult to make a nondemolition measurement.
Nevertheless, we have shown, for particular cases, that this measurement is quite realizable, although
 only for a restricted range of the parameters. For performing this kind of measurement, it is
important to take into account the fact that shortening the measurement time
requires a growth of the leakage, $\kappa$, thus decreasing the resolution of
the resonator lines and decreasing the quality factor. Moreover, a
very strong drive can violate the conditions required for the dispersive
measurement and it can shorten the lifetime of the qubit in the excited
state. On the other hand, using resonators and qubits with high
frequencies enables the choice of suitable parameters.

In this paper, we have only touched the problems of the qubit-resonator-drive noises that are very important for quantum measurements. The measurement induced
noise arises from the second and third terms in square brackets of
Eq. (\ref{tve}). They can be accounted for straightforwardly.
Nevertheless, the complete solution of this problem is possible only by
accounting for the measuring scheme and the intrinsic noises of the measuring
device, and by taking into consideration the noise produced by the amplifiers. All these issues  require further consideration.

\section{Acknowledgment}

We thank D. Kinion  for useful discussions.  This work was carried out under the auspices of the National Nuclear Security Administration of the
U.S. Department of Energy at the Los Alamos National Laboratory under Contract No. DE-AC52-
06NA25396, and was funded by the Office of the Director of National Intelligence (ODNI), and Intelligence Advanced Research Projects Activity (IARPA). All statements of fact, opinion or conclusions contained herein are those of the authors and should not be construed as representing the official views or policies of IARPA, the ODNI, or the U.S. Government.

\newpage \parindent 0 cm \parskip=5mm

\end{document}